\documentclass[aps,longbibliography,showpacs,twocolumn,superscriptaddress]{revtex4-1}
\usepackage{amsmath,amssymb,amsfonts,bm}
\usepackage{graphicx}
\usepackage{epstopdf}
\usepackage{dcolumn}
\usepackage{mathrsfs}
\usepackage{bbold}
\usepackage{dsfont}
\usepackage{float}
\usepackage[colorlinks=true,linkcolor=blue,citecolor=blue, urlcolor=blue]{hyperref}
\usepackage{tgtermes}
\usepackage{multirow}
\usepackage{bbm}
\usepackage{wasysym}
\usepackage{newtxtext}
\usepackage[varvw]{newtxmath}
\usepackage{hyperref}
\usepackage{array}
\usepackage{bm}
\usepackage{color}
\usepackage{float}
\usepackage{graphicx}
\usepackage{natbib}
\usepackage{newtxtext}
\usepackage{newtxmath}
\usepackage[caption=false]{subfig}
\usepackage{mathrsfs}
\usepackage{amsmath,amssymb}
\usepackage{verbatim}
\usepackage{amsfonts}
\usepackage{dcolumn}
\usepackage{ulem}

\newcommand{\beq}{\begin{eqnarray} }
\newcommand{\eeq}{\end{eqnarray} }
\newcommand{\Beq}{\begin{eqnarray*} }
\newcommand{\Eeq}{\end{eqnarray*} }
\newcommand{\Bmat}{\left(\begin{matrix}}
\newcommand{\Emat}{\end{matrix}\right)}
\newcommand{\up}{\uparrow}
\newcommand{\dn}{\downarrow}

\newcommand{\bit}{\begin{itemize} }
\newcommand{\eit}{\end{itemize} }
\newcommand{\ben}{\begin{enumerate} }
\newcommand{\een}{\end{enumerate} }

\begin{document}

\title{Excitonic Quantum Anomalous Hall Effect in Collinear Magnets Without Spin-Orbit Coupling}

\author{Xingxing Liu}
\email{These two authors contribute equally to this work.}
\affiliation{Department of Physics and Beijing Key Laboratory of Opto-electronic Functional Materials and Micro-nano Devices, Renmin University of China, Beijing, 100872, China}
\affiliation{Key Laboratory of Quantum State Construction and Manipulation (Ministry of Education), Renmin University of China, Beijing, 100872, China}
\author{ChaoYang Tan}
\email{These two authors contribute equally to this work.}
\affiliation{Department of Physics and Beijing Key Laboratory of Opto-electronic Functional Materials and Micro-nano Devices, Renmin University of China, Beijing, 100872, China}
\affiliation{Key Laboratory of Quantum State Construction and Manipulation (Ministry of Education), Renmin University of China, Beijing, 100872, China}

\author{Peng-Jie Guo}
\email{guopengjie@ruc.edu.cn}
\affiliation{Department of Physics and Beijing Key Laboratory of Opto-electronic Functional Materials and Micro-nano Devices, Renmin University of China, Beijing, 100872, China}
\affiliation{Key Laboratory of Quantum State Construction and Manipulation (Ministry of Education), Renmin University of China, Beijing, 100872, China}

\author{Zhong-Yi Lu}
\email{zlu@ruc.edu.cn}
\affiliation{Department of Physics and Beijing Key Laboratory of Opto-electronic Functional Materials and Micro-nano Devices, Renmin University of China, Beijing, 100872, China}
\affiliation{Key Laboratory of Quantum State Construction and Manipulation (Ministry of Education), Renmin University of China, Beijing, 100872, China}

\author{Zheng-Xin Liu}
\email{liuzxphys@ruc.edu.cn}
\affiliation{Department of Physics and Beijing Key Laboratory of Opto-electronic Functional Materials and Micro-nano Devices, Renmin University of China, Beijing, 100872, China}
\affiliation{Key Laboratory of Quantum State Construction and Manipulation (Ministry of Education), Renmin University of China, Beijing, 100872, China}

\date{\today}

\begin{abstract}
Spin–orbit coupling (SOC) is thought to be necessary in realizing quantum anomalous Hall (QAH) insulators in magnetic materials. In this Letter, we propose an exciton-condensation mechanism to realize QAH effect in collinear magnets with negligible spin-orbit coupling. This mechanism is realized by two steps: first prepare a spin-splitting nodal-ring band structure, and then gap out the nodal-ring via triplet exciton condensation. A nonzero Chern number can be obtained if the in-plane spin texture resulting from the triplet exciton condensation is noncollinear in momentum space. We show that the electron-phonon coupling can switch the spin texture from a colinear pattern to a noncolinear one and plays an essential role in realizing QAH effect. The above mechanism is not only suitable for ferrogmagnets but also applicable for altermagnets. Finally,  through first-principles calculations we propose the bilayer material $\mathrm{V_2SeTeO}$ to be a promising candidate of excitonic QAH insulator. 

\end{abstract}

\maketitle

{\it Introduction.} 
Quantum anomalous Hall (QAH) insulators are topological phases in 2-dimensions (2D) which possess a bulk gap and chiral edge states without applying external magnetic fields. The Hall conductance of such a state is quantized to an integer $C$ times $e^2/h$ at temperatures below the bulk gap. The topological invariant of a QAH insulator is characterized by the Chern number $C$, which is proportional to the integration of Berry curvature for occupied states over the Brillouin zone (BZ). Since the Chern number reverses its sign under time-reversal operation $T$, QAH effect does not appear in systems having anti-unitary symmetries like $T$, $IT$, $(T|\tau)$ or other types of effective time reversal symmetries, where $I$ and $\tau$ represent space-inversion and fractional translation operations, respectively.  QAH insulators were realized in theoretical models of spinless fermions containing staggered flux\cite{Haldane} or ferromagnetic (FM) electron systems having finite SOC\cite{QiWuZhang}. Experimentally realized QAH insulators all belong to FM systems\cite{LIuZhangQi_ARCMP16, Xue_Science13, WuShanYan_PRL14} where SOC plays an essential role in generating Berry curvature and hence the nonzero Chern number. 

Recently, an emerging magnetic system called altermagnetism \cite{altermagnetism-PRX1,PRX2,PRX3} was shown to contain spin-splitting electronic band structure. Unlike FM systems, altermagnetic (AM) materials exhibit $k$-dependent spin polarizationan which are categorized into $d$-wave, $g$-wave, $i$-wave and so on. AM have many interesting consequences, including pizomagnetic effect\cite{LiuJunWei, LiuJunWei2, Piezomagnetization1, Piezomagnetization2, Piezomagnetization3, Piezomagnetization4, Piezomagnetization5}, giant magnetic resistance\cite{PRX3}, spin current\cite{LiuJunWei, LiuJunWei2}, and so on. Especially, due to the spin splitting spectrum, altermagetism  can realize anomalous Hall effect\cite{AH1, AH2, AH3, AH4, AH5, AH6} or even QAH effect\cite{GuoLiuLu_npj23} in the presence of finite SOC. On the other hand, when SOC is negligible, AM have enhanced symmetries described by collinear spin space groups\cite{SSG1, SSG2, SSG3}, which contain spin-only operations $\left(C_{2}^{x}T||T\right)$, $\left(C_{\infty}^{z}||E\right)$ (assuming that the magnetic momentum are polarizing along $z$-direction). The spin space groups can support many interesting physical phenomena, such as topological semimetals with new symmetry invariants\cite{YangLiuFang21B, YangLiuFang24NC, SM1, SM2, SM3, SM4}, or topological insulators/superconductors\cite{TI1,TI2,TI3, SongZD25L, TSC1,TSC2,TSC3} that can not be realized in Magnetic space groups in which SOC is significant. However, due to the effective time reversal symmetry $\left(C_{2}^{x}T||T\right)$, the total Chern number in an insulating altermagnet must be zero. Therefore, in order to realize QAH effect in such systems, a mechanism is needed to spontaneously break the effective time reversal
symmetry via certain interactions. 

Actually, exciton condensation provides a possible route to break the
effective time reversal symmetry. Like Cooper pairs in superconductors, excitons can condense \cite{EC1, EC2} under circumstances where the energy cost in forming an electron-hole pair becomes negative. 
The spontaneous condensation of excitons can induce a phase transition from a metal into an insulator \cite{MacDonald_PRB, Metal2}.
For magnetic systems, the excitons may be either spin-singlets or spin-triplets, providing more possibilities in the resulting phases of matter\cite{TEI1, TEI2, spintex, spintriplet}. Especially, like topological $(p_x+ip_y)$-wave superconductors, the condensing of triplet exciton can also give rise to a topological state. 

In this Letter, we demonstrate the mechanism to realize excitonic QAH insulator by constructing microscopic lattice models in collinear magnets. While screened Coulomb interaction can give rise to triplet exciton condensation with collinear in-plane spin texture which preserves an effective time reversal symmetry $\tilde T\equiv \left(C_{2}^{\pmb n}T||T\right)$ ($\pmb n$ is a unit vector perpendicular to the direction of the magnetic momentum), electron-phonon interactions can break $\tilde T$ by switching the in-plane spin texture to a non-collinear one, resulting in a QAH insulator with nonzero Chern number. We first illustrate such a mechanism via a ferromagnetic model, and then show that the same mechanism still works for altermagnets.  Furthermore, via the first-principles calculations, we propose a candidate material -- bilayer $\mathrm{V_2SeTeO}$\cite{VSeTe} to promisingly realize the excitonic QAH effect under charge-charge interactions. 

\begin{figure}[t]
	\includegraphics[width=8.5cm]{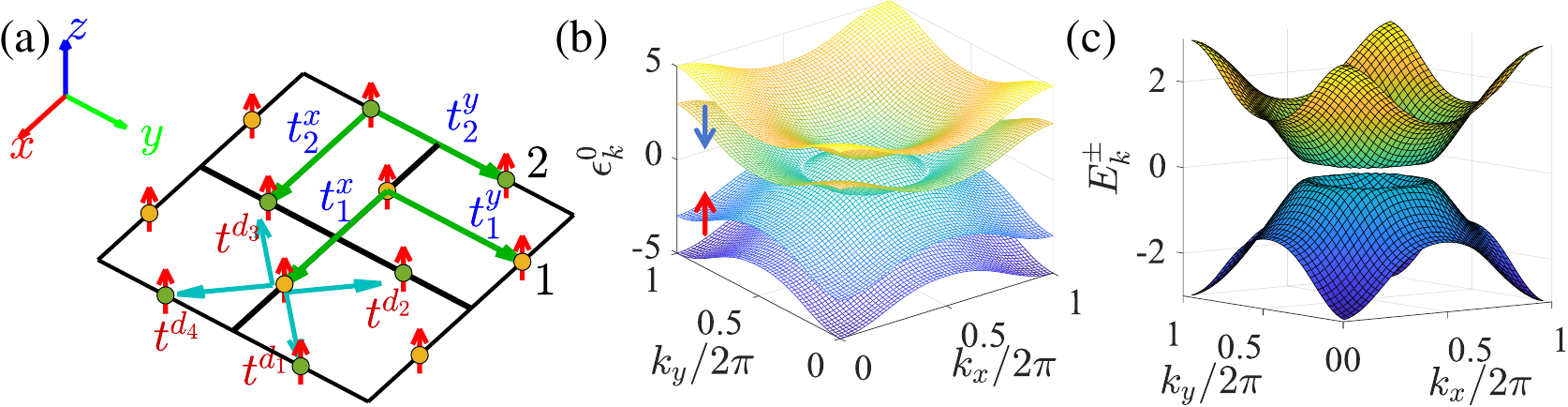}
	\caption{ Model for the Ferromagnetism. (a) Illustration of the Hamiltonian in Eq.(\ref{Eq1-FM}). (b) Four-band spin-splitting energy dispersion 
	with $t^{d}=1$, $t=0.6$, $m=2$. The red arrow indicates that the spin polarization of the top valence band is spin up, while the blue arrow represents one of the bottom conducting band is spin down. (c) Excitation energies $ E_{\pmb{k}}^{\pm} $ of excitonic insulators being gapped.}
	\label{lattice-FM}
\end{figure}

{\it Ferromagnetic Model with a nodal ring.} We first construct a ferromagnetic model on square lattice with nodal-ring structure. Such a structure requires at least two sublattices, therefore, we consider an FM model with two sites in each unit cell labeled by sublattice index $ \alpha = 1,2$, where each site carries uniform magnetic moment $\pmb m_1=\pmb m_2=m\hat{\pmb z}$. Without loss of generality, we assume that the easy axis of the magnetic momentum is parallel to $\hat{\pmb z}$. Supposing that each site contributes one itinerary electron, up to next-nearest neighbor terms the tight-binding model of the electrons on the lattice reads (see Fig.\ref{lattice-FM}(a) for illustration), 
\beq\label{Eq1-FM}
H_0^{\rm fm}&\!=\!& \sum_j \Big( \sum_{d_i}t^{d_i}C_{1,j}^\dag C_{2,j+\pmb{d}_i}  
+\!\! \sum_{\alpha, p=x,y}  t_\alpha^pC_{\alpha,j}^\dag C_{\alpha,j \pm \pmb p} \nonumber \\ 
& &+\text{h.c.} \Big) 
+ \sum_{\alpha,j}C_{\alpha,j}^\dag \big( \mu_\alpha \sigma_0+\pmb{m} \cdot{\pmb{\sigma}\over2}\big) C_{\alpha,j}
\eeq
where $C_{\alpha, j}^\dag=(c_{\alpha, j\uparrow}^\dag, c_{\alpha, j\downarrow}^\dag) $ are electron creation operators, $ t^{d_i}, t_\alpha^{p} $ are the nearest neighbor and next-nearest neighbor hopping coefficients with the lattice vectors given by $ \pmb{x}=a\hat{\pmb {x}}$, $\pmb{y}=a\hat{\pmb{y}}$ (with $a$ the lattice constant and $\hat{\pmb{x}} $ and $ \hat{\pmb{y}} $ the unit vectors) and $ \pmb{d}_{1}=\pmb{x}+\pmb{y}$, $ \pmb{d}_{2}=\pmb y$, $\pmb{d}_3=0$, $\pmb{d}_4=\pmb x$. The term $ \pmb{m}_{\alpha} \cdot {\pmb{\sigma}\over2} $ denotes the Zeeman coupling between the electron spin and the static collinear magnetic moment, and $ \mu_{\alpha}$  (with $\sigma_0$ standing for identity matrix) represents the chemical potential term on the $\alpha $-sublattice and has been fixed to zero. For simplicity, we set $t^{d_1}=t^{d_2}=t^{d_3}=t^{d_4}=t^d$ and $t_1^{x}=-t_1^{y}=-t_2^{x}=t_2^{y}=t$ in later discussion. Spin-orbit coupling has been ignored here. 

After the Fourier transformation, the Hamiltonian [Eq.(\ref{Eq1-FM})] can be rewritten in momentum space as 
$$H=\sum_k \left( C_{1k}^\dag \Gamma_k^{12} C_{2k}+\rm{h.c.} \right)+\sum_{\alpha,k} C_{\alpha k}^\dag \Gamma_k^\alpha C_{\alpha k},
$$
where $C_{\alpha k}^\dag=(c_{\alpha k\up}^\dag,c_{\alpha k\dn}^\dag)$, $\Gamma_k^{12}=t^d(1+e^{i k_x}+e^{i k_y}+e^{i(k_x+k_y)})$ stand for the intersublattice terms and $\Gamma_k^\alpha=(-1)^\alpha2t\left( \cos k_x-\cos k_y\right) +\left( \mu_\alpha+\pmb{m} \cdot {\pmb{\sigma}\over2} \right)$ are the intrasublattice terms. 
The FM system has a spin point group symmetry generated by $ \left(E|| C_{4}^{z}\right)$, 
$ \left(E||C_{2}^{x}\right) $, $ \left(E||C_{i}\right) $, $ \left(C_{2}^{x}T||T\right) $, $ \left(C_{\infty}^{z}||E\right) $ where the notation $ \left( s||l\right)  $ denotes a combined operation of the spin operation $s$ and the lattice operation $l$. Here $\{\left( C_{\infty}^{z}||E\right)\}$ form a $U(1)$ spin rotation symmetry group and yield the conservation of $S_z$ for the electrons. 
The energy spectrum (referred as $ \epsilon_{d\pmb{k}}^{0}$ later) of the above Hamiltonian is shown in Fig.\ref{lattice-FM}(b), where the conducting band cross the valence band with a nodal ring at the fermi level. Since the valence and conducting bands carry opposite spin, as shown in Fig.\ref{lattice-FM}(b), the nodal ring is protected by the $U(1)$ symmetry. Such a nodal-ring structure supports triplet excitons.  
 
{\it Column interaction and Triplet Exciton condensation.}
 The attractive interaction between the particle and hole excitations can trigger the exciton condensation. Hence we consider the electron-electron repulsive Column interaction of the form $H_c = \sum_r {1\over r} \psi_i^\dag \psi_{i+r}^\dag \psi_{i+r}\psi_i$. In momentum space, the Column interaction is given by
$ H_c = \frac{1}{2N}\sum_{\pmb{k},\pmb{k}^{'},\pmb{q}} \sum_{\beta,\beta^{'}}U_c(\pmb{q})c_{\beta,\pmb{k}+\pmb{q}}^\dag c_{\beta^{'},\pmb{k}^{'}-\pmb{q},}^\dag c_{\beta^{'},\pmb{k}^{'}} c_{\beta,\pmb{k}},$
where $N$ is the number of primitive cells and $\beta,\beta^{'}$ are combined sublattice and spin indices. 
Considering the screening (here we adopt the dual-gate screened form in 2D continuum limit \cite{Column,Column2}), the interaction potential is approximated as $U_c(\pmb{q})= \frac{g}{\left| \pmb{q}\right|} \tanh(\xi\left| \pmb{q}\right|)$, with $\xi$ the gate distance which has been fixed to $\xi=10$ in later discussion, and $g$ the interaction strength which is considered as an adjustable interaction parameter. 
On the other hand, due to electron-phonon coupling, one also needs to consider an effective attractive interaction $U_p(\pmb{q})= -V_{\rm ph}$ between the electrons within the Debye energy. Then the total interaction is given by 
$$H_I = \frac{1}{2N}\sum_{\pmb{k},\pmb{k}^{'},\pmb{q}} \sum_{\beta,\beta^{'}}U(\pmb{q})c_{\beta,\pmb{k}+\pmb{q}}^\dag c_{\beta^{'},\pmb{k}^{'}-\pmb{q},}^\dag c_{\beta^{'},\pmb{k}^{'}} c_{\beta,\pmb{k}},$$
 with
$
U(\pmb{q})=U_c(\pmb{q})+U_p(\pmb{q}) =\frac{g}{\left| \pmb{q}\right| } \tanh(\xi\left| \pmb{q}\right|)-V_{\rm ph}.
$
  
\begin{figure}[t]
	\includegraphics[width=8cm]{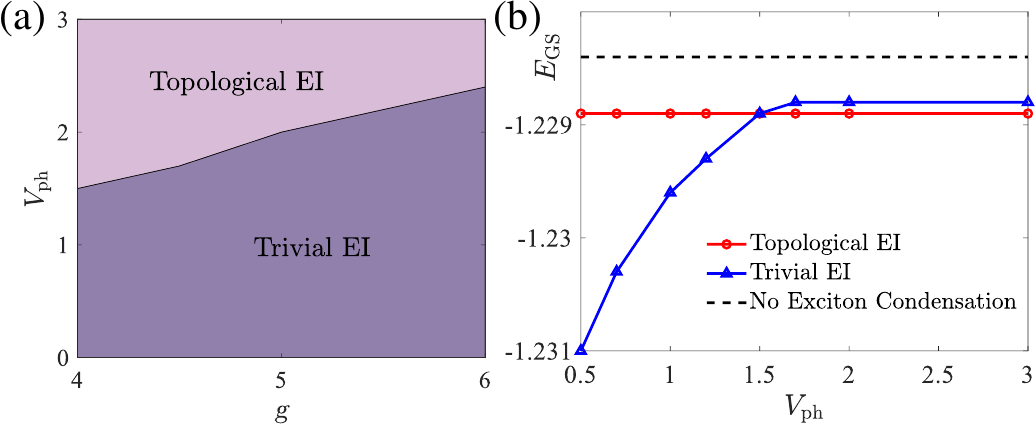}
	\caption{ (a) The phase diagram based on the mean-field calculation in different parameters $g$ and $V_{\rm ph}$ for FM model. (b) The total ground
		energy $E_{\rm GS}$ of topological EIs, trivial EIs and metals (no exciton condensation) with various $V_{\rm ph}$ with $g=4$.}
	\label{fig:Transition}
\end{figure}
 
In the following we study the consequence of the interactions using self-consistent mean field theory. We first project the interactions onto the two bands close to the fermi level, namely the top valence (TV) band and bottom conducting (BC) band\cite{Column}. By introducing the band-projected density operator $\rho_{\pmb{q}}=\sum_{d,d^{'},\pmb{k}} F_{\pmb{k}-\pmb{q},\pmb{k}}^{dd^{'}} b_{d,\pmb{k}-\pmb{q}}^\dag b_{d^{'}\pmb{k}}$ with $b^\dag_{d\pmb k} =\sum_\beta c^\dag_{\beta \pmb k} \varphi_{\beta, d\pmb k} $ the eigenmode creation operator and 
$F_{\pmb{k},\pmb{k}^{'}}^{dd^{'}}
=\sum_\beta \varphi_{\beta,d\pmb k}^*\varphi_{\beta, d'\pmb k'}$ the form factor, the projected total Hamiltonian reads
\begin{eqnarray}\label{Eq4}
	H&=& \sum_{d,\pmb{k}} \epsilon_{d\pmb{k}}^{0} b_{d\pmb{k}}^\dag b_{d\pmb{k}}
	+ \frac{1}{2N}\sum_{\pmb{q}} U(\pmb{q}) \rho_{\pmb{q}}^\dag \rho_{\pmb{q}}.
\end{eqnarray}
Here the band index $d,d^{'}=v,c$ with $v,c$ standing for the TV and BC band, respectively. Notice that the spin index is included in the band index $d,d'$ due to the spin polarization. Introducing the exciton condensate parameter $\Delta_{\pmb{k}}$, the Hartree-Fock mean-field Hamiltonian can be obtained 
\begin{eqnarray}\label{Eq:MF}
	\!\!\!\!\! H_{\rm MF}=\! E_{0}+\!\!\sum_{\pmb{k},d}  \epsilon_{d\pmb{k}}^{0} b_{d\pmb{k}}^\dag b_{d\pmb{k}} 
	-\!\!\sum_{\pmb{k}} ( \Delta_{\pmb{k}}^{*} b_{c\pmb{k}}^\dag b_{v\pmb{k}}+\text{h.c.})
\end{eqnarray}
with energy spectrum
$E_{\pmb{k}}^{\pm}={1\over2}(\epsilon_{c\pmb{k}}^0+\epsilon_{v\pmb{k}}^0) \pm \sqrt{\epsilon_{\pmb{k}}^{2}+|\Delta_{\pmb{k}}|^{2}}$  where $\epsilon_{\pmb{k}}={1\over2}(\epsilon_{c\pmb{k}}^0-\epsilon_{v\pmb{k}}^0)$. 
Due to the nodal line structure, we only considered the excitons carrying zero total momentum $Q=0$. 

\begin{figure}[t]
	\includegraphics[width=8.5cm]{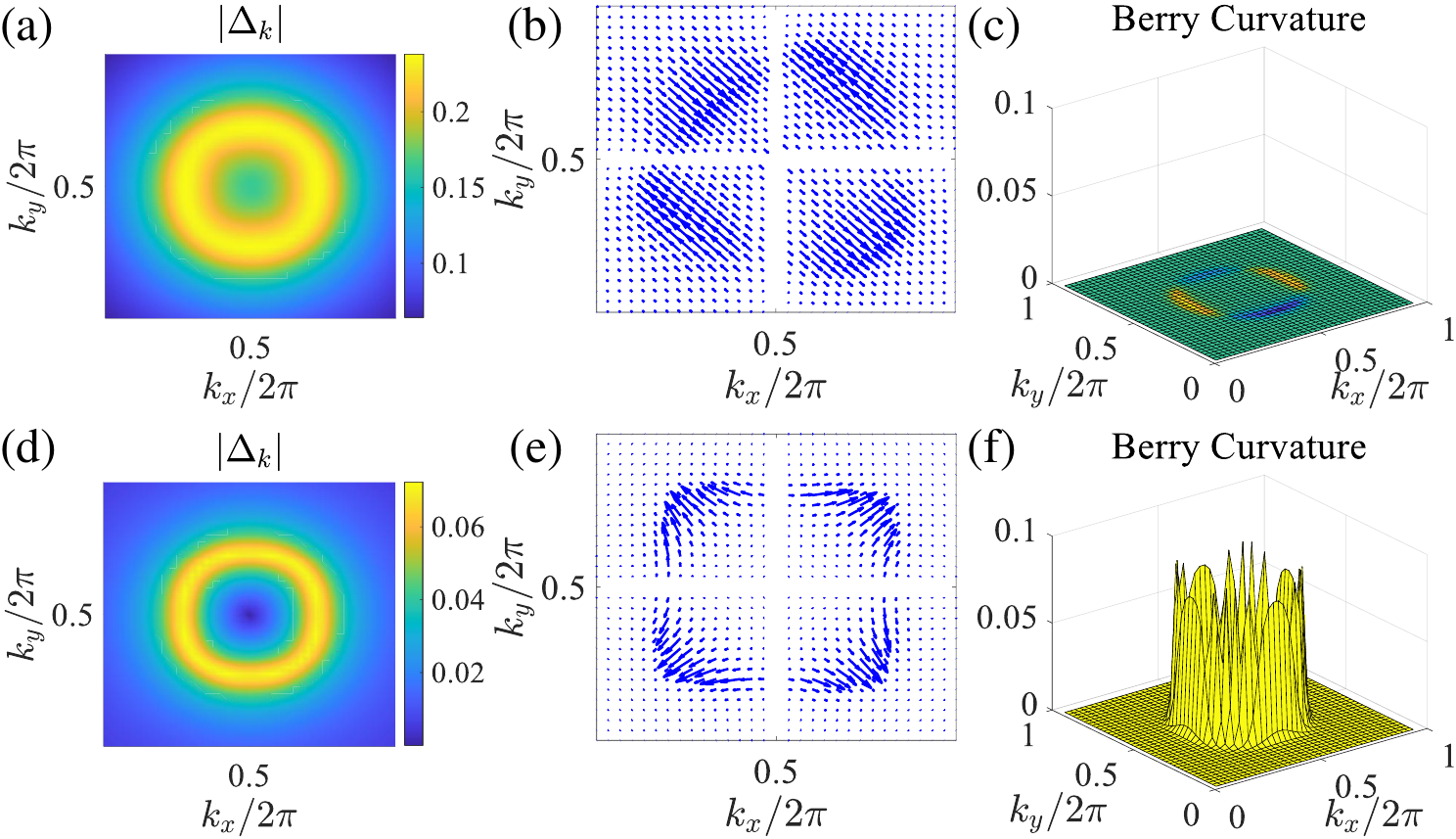}
	\caption{ The results of exciton condensation in ferromagnetism for $g=4$. (a) Magnitude of the order parameters $|\Delta_{\pmb{k}}|$ for $V_{\rm ph}=0$, (b) the spontaneous $d$-wave collinear spin textures and (c) the berry curvature with $C=0$. While, (d) magnitude of the order parameters $|\Delta_{\pmb{k}}|$ for $V_{\rm ph}=3$, (e) the spontaneous $(p_x+i p_y)$-wave non-collinear spin textures and (f) the berry curvature with $C=1$.} \label{fig:FM-EC}
\end{figure}

At zero temperature, the parameters  $\Delta_{\pmb{k}}$ are determined self-consistently by the gap equations (see End Matter).
Like conventional superconductors, the energy spectrum of exciton condensate generally opens a gap and form an excitonic insulator (EI) [Fig.\ref{lattice-FM}(c)] due to the mixing of the conducting band and valence band by the condensate $\Delta_{\pmb{k}}$ [Fig.\ref{fig:FM-EC}(a)\&(d)]. However, unlike the Cooper pairs in superconductors, the `pairing' symmetry between particles and holes is unrelated to the total spin of the exciton-pair. Hence $s$-wave, $p$-wave, $d$-wave, $g$-wave and other pairing-wave symmetries are all possible solutions for triplet exciton condensate.

The condensing of triplet excitons spontaneously breaks the $U(1)$ spin rotation symmetry, and may cause in-plane magnetization of the electron spin. Therefore, we can use the spin textures (in-plane magnetic moments of electrons at momentum points in the BZ) to define the `pairing symmetry' of the triplet exciton condensation (see End Matter).  For instance, in an $s$-wave exciton condensate, the electrons (in the conducting and valence bands) exhibit nonzero in-plane magnetization, $M_{x,y}=  \frac{\mathscr g\mu_B}{N}\sum_{i, \alpha} \langle  {\rm GS} | S^{x,y}_{i\alpha}|{\rm GS}\rangle = \frac{\mathscr g\mu_B}{N}\sum_{\pmb k} \langle S^{x,y}(\pmb k)\rangle$, where $\mathscr g$ is the Lander g-factor and $\mu_B$ is the Bohr magnetron,
$|{\rm GS}\rangle$ is the ground state of triplet exciton condensate, and $\langle S^{x,y}(\pmb k)\rangle$ are defined in End Matter.  The in-plane magnetization $(M_{x},M_{y})$ is an observable quantity and can be adopted as order parameters of the triplet exciton condensate. 

{\it Excitonic QAH insulators in Ferromagnetism.}
When $V_{\rm ph}=0$, the $d$-wave exciton condensate is found to be the ground state. The in-plane magnetization of $(M_x, M_y)$ shows a $d$-wave texture in the BZ with zero net in-plane magnetization. Substituting $\Delta_{\pmb{k}}$ into (\ref{Eq:MF}), one obtains $\tilde T H_{\rm MF}^{(d)}\tilde T^{-1}=H_{\rm MF}^{(d)}$, meaning that the $d$-wave excitonic insulator (EI) does not break $\tilde T$ symmetry. As a consequence, the $d$-wave EI carries trivial Chern number [see Fig.\ref{fig:FM-EC}(b)\&(c)].

Actually, there indeed exist a solution with nontrivial Chern number $C=\pm1$ in which the triplet excitons condense in a $(p_x\pm ip_y)$-wave like pattern with noncollinear spin textures. The textures of in-plane magnetization $\langle S^x(\pmb k)$, $\langle S^y(\pmb k)\rangle$ are both of $p$-wave. Since the $\langle S^x(\pmb k)\rangle$ and $\langle S^y(\pmb k)\rangle$ are contributed by the real and imaginary parts of the order parameters $\Delta_{\pmb k}$ respectively, this condensate can be regarded as of $(p_x\pm ip_y)$-wave symmetry. The effective time reversal symmetry $\tilde T$ is spontaneously broken since the solution with $C=1$ is transformed in the one with $C=-1$ by $\tilde T$ and vice versa. Unfortuanately, the solutions with nonzero Chern number are energetically higher than the $d$-wave solution (see Fig.\ref{fig:Transition}).  

However, by considering electron-phonon coupling and tuning the interaction parameters $V_{\rm ph}$ and $g$, the ground state can be switched into the solution with $(p_x+ip_y)$-wave excitonic paring and nonzero Chern number. Hence, an exitonic QAH insulator is realized without using spin-orbit coupling. The excitonic QAH insulators with nonzero Chern numbers $C\neq0$ appear at the relatively large $V_{\rm ph}$(see Fig.\ref{fig:Transition}). 
The magnitude of order parameters, the spin texture in momentum space, and the Berry curvature of the $C=1$ solution (with $V_{\rm ph}/g$=0.75) are shown in Fig.\ref{fig:FM-EC}(d)$\sim$(f), respectively. The total in-plane magnetization still vanishes, but the nontrivial Chern number indicates that it has chiral edge states and quantized Hall conductance at low temperatures. To verify this, we consider a system with periodic boundary condition along the $x$-direction and open boundary condition along the $y$-direction. Two branches of gapless chiral edge states are found on the open boundaries, as shown in the End Matter. Since the Coulomb interaction is long-ranged, the Hartree-Fock mean field Hamiltonian contains long-range hopping terms. For this reason, the edge states is not extremely localized on the edge. Instead, the edge state has a relative large width (see Fig.\ref{Wf} in End Matter) comparing to the edge states in QAH insulators caused by SOC\cite{GuoLiuLu_npj23}.  The gapless chiral edge modes indicate that the Hall conductance is quantized when the fermi energy falls in the bulk gap.

The appearance of the topological EI phase can be qualitatively interpreted as the following. The electron-phonon coupling gives rise to an effective repulsion between electrons and holes close to the fermi surface hence increases the energy of the exciton condensates. Since the order parameter $\Delta_{\pmb k}$ of the topological EI contains one node at the $(\pi,\pi)$ point but the trivial EI has no nodes (see Fig.\ref{fig:FM-EC}(a)\&(d)), the speed of the energy increasing of the topological state due to the electron-phonon coupling is slower than that of the trivial state (actually the energy of the topological EI is independent on $V_{\rm ph}$ because the $V_{\rm ph}$ term vanishes when projected to the $(p_x+ip_y)$-wave channel).  As a result, a phase transition to the topological EI can be realized with the increasing of  $V_{\rm ph}$ (see Fig.\ref{fig:Transition}).

\begin{figure}[h]
	\includegraphics[height=4.3cm]{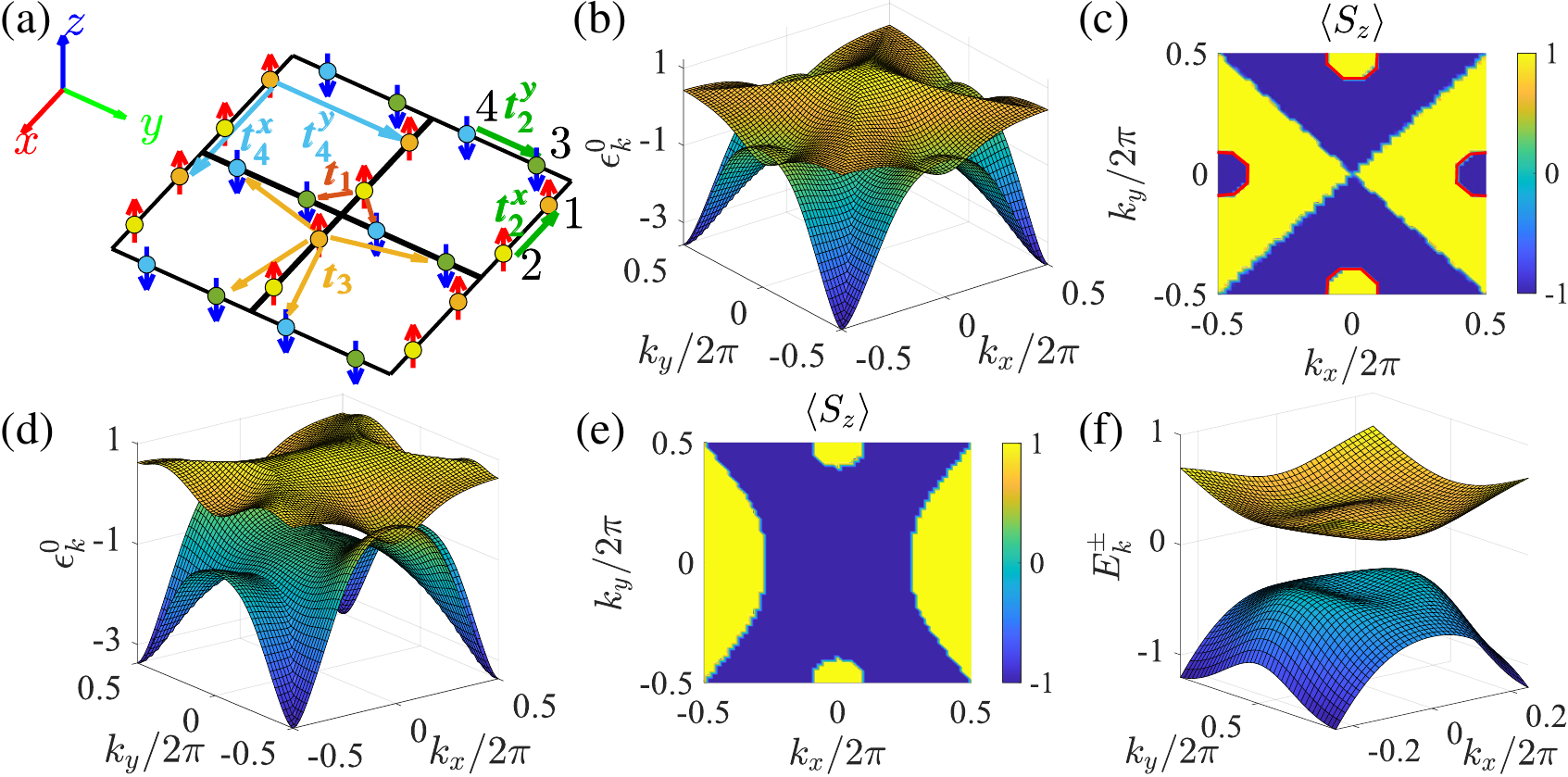}
	\caption{ Model for the altermagnetism with a $d$-wave symmetry. (a) Illustration of the Hamiltonian in Eq.(\ref{AMEq1}). (b) The two middle ones in the eight-band spin-splitting energy dispersion diagram of Hamiltonian in Eq.(\ref{AMEq1}) with $t_{1}=0.5$, $t_{2}^{x,y}=1.2$, $t_{3}=0.7$, $t_{4}^{x,y}=0.4$ and $m=4$. (c) The spin-polarization of the top valence band, which is opposite to that of the bottom conducting band. The red solid lines stands for the nodal rings. (d) The energy dispersion, (e) the spin-polarization and (f) excitation energies $ E_{\pmb{k}}^{\pm} $ of excitonic insulators after applying applying strain along the $y$-axis.}
	\label{lattice-AM}
\end{figure}

{\it Excitonic QAH insulators in altermagnetism.} We now switch to realize excitonic QAH insulators in altermagnetic models using a similar mechanism. 
To obtain a spin splitting nodal-ring structure similar to Fig.\ref{lattice-FM}(b), one needs at least four sites in each unit cell. We label the sublattice index as $ \alpha = 1,2,3,4$, where the sites 1,2 carry positive magnetic moment $\pmb m_1=\pmb m_2=m\hat{\pmb z}$ and 3,4 sites carry negative magnetic moment $\pmb m_3=\pmb m_4=-m\hat{\pmb z}$. The four sites are related to each other by the anti-unitary spin point group symmetry $(T||C^z_4T)$, where $C^z_4$ stands for 4-fold lattice rotation (see Fig.\ref{lattice-AM}(a) for illustration). The tight-binding model of the electrons on the lattice reads (see Fig.\ref{lattice-AM}(a) for illustration) in End Matter.
After the Fourier transformation, the Hamiltonian can be rewritten in momentum space as $H_0=\sum_k\psi_k^\dag \mathscr H_k \psi_k$ in the bases $\psi_k^\dag =[c_{1k\up}^\dag,c_{1k\dn}^\dag, c_{2k\up}^\dag, c_{2k\dn}^\dag, c_{3k\up}^\dag, c_{3k\dn}^\dag, c_{4k\up}^\dag,c_{4k\dn}^\dag]$, where
\begin{eqnarray}\label{AMEq2}
	\mathscr H_k = 
	\Bmat 
	\Gamma_k^{(1)} & \Gamma_k^{(21)*} &\eta_k^*\sigma_0 &\chi_k^*\sigma_0\\
	\Gamma_k^{(21)}  &\Gamma_k^{(2)}  &\chi_k\sigma_0 &\eta_k\sigma_0\\	
	\eta_k\sigma_0&	\chi_k^*\sigma_0 &\Gamma_k^{(3)} &\Gamma_k^{(43)*} \\		 
	\chi_k\sigma_0 &\eta_k^*\sigma_0 &\Gamma_k^{(43)}&\Gamma_k^{(4)}
	\Emat	
\end{eqnarray}
is the Hamiltonian matrix at $k$ point with $\Gamma_k^{(\alpha)}=\pmb m_\alpha \cdot {\pmb \sigma\over2}+\left( 2t_4^x \cos k_x+2t_4^y \cos k_y \right)\sigma_{0}$, $\Gamma_k^{(21)}=t_2^x  \left( 1+e^{-ik_x} \right) \sigma_{0}$, $\Gamma_k^{(43)}=t_{2}^y \left( 1+e^{ik_y} \right) \sigma_{0}$, $\eta_k= t_{1}+t_3\left( e^{-ik_x}+e^{-ik_y} \right) $ and $\chi_k= t_{1}+t_3\left( e^{-ik_x}+e^{ik_y} \right) $.
The parameters $ t_{1}$ and $t_3 $ are the nearest neighbor and next-nearest neighbor hopping coefficients between the cites carrying opposite magnetic moment, while $t_2^{x,y}$ and $t_4^{x,y}$ are those between the cites with the same magnetic moment (see the End Matter for the Hamiltonian in real space). The term $\pmb{m}_{\alpha} \cdot {\pmb{\sigma}\over2}$  denotes the Zeeman coupling and $ \mu_{\alpha}$ represents the chemical potential on the $\alpha$-sublattice which has been fixed to a constant $\mu_\alpha=0.03$. The AM system has a spin point group symmetry generated by $\left(T || TC_{4}^{z}\right)$,  $ \left(E||C_{2}^{x}\right) $, $ \left(E||C_{i}\right) $, $ \left(C_{2}^{x}T||T\right) $ and $\left(C_{\infty}^{z}||E\right)$. The  symmetry $(T||C^z_4T)$ ensures the cancelation of net magnetic moment and the $d$-wave symmetric spin-splitting band structure (see Fig.\ref{lattice-AM}(b) for the two bands close to the Fermi level). 

As shown in Fig.\ref{lattice-AM}(c), there are two nodal rings locating at X and Y points  
which are related by 
the $\left(T || TC_{4}^{z}\right)$ symmetry.  
Since the TV and BC bands also have opposite spin polarization, triplet excitons are also supported.  
Following Ref.\cite{GuoLiuLu_npj23}, we apply uniaxial strain to remove one of the nodal rings
such that the band structure is similar to the FM model. When the strain is applied along the $y$-axis such that $t_2^x > t_2^y$ and $t_4^x > t_4^y$, then the nodal ring at X point shrinks  its size and eventually gaps out, leaving only the ring at Y point untouched [see Fig\ref{lattice-AM}(d)\&(e)]; similarly one can also gap out the ring at the Y point and only keep the one at X point by applying strain along the $x$-axis. 
The final band structure is similar to the FM model except that the the nodal ring is now centered at X (or Y) point instead of the $\Gamma$ point. The solutions of the self-consistent gap equations in the momentum area enclosing the nodal line also contain a trivial triplet-exciton insulator and a topological exciton insulator (see End Matter for details).

Since the center of the nodal ring is not locating at the $\Gamma$ point, the `pairing symmetry' of the triplet exciton insulators are also different from that in the FM model. At the small $V_{\rm ph}$ side, the trivial EI is of $p$-wave and the Chern number is $C=0$ due to the unbroken $\tilde T$ symmetry. The texture of the in-plane spin texture shows a $p$-wave collinear pattern (see Fig.\ref{AM-EC}(c) in End Matter) with vanishing net in-plane magnetization. However, in the excitonic QAH insulator (with Chern number $|C|=1$) at the large $V_{\rm ph}$ side, the texture of the in-plane magnetization exhibits a $(s+id)$-wave noncollinear pattern (see Fig.\ref{AM-EC}(g) in End Matter). Beside the nontrivial topological properties (including chiral edge states and quantized Hall conductance at low temperatures), the QAH insulator also exibits net in-plane magnetization of order $\sqrt{M_x^2+M_y^2} \sim 0.02\mu_B$ per site (with $V_{\rm ph}=1.3$ and $g=1$), which is an experimentally observable quantity. Notably, the detection of the in-plane net magnetization can be considered as a directly evidence of the condensation of the triplet excitons. 

Here we summarize several conditions which may aid the formation of excitonic QAH insulators:
\ben
\item[(1)] existence of spin-$U(1)$ symmetry protected nodal-lines (odd number of rings preferred) with spin splitting;
\item[(2)] significant intensity of electron-phonon coupling; 
\item[(3)] conducting \& valence bands being from different layers. 
\een
Here condition (1) is imposed to facilitate the condensation of triplet excitons under Coulomb interactions, especially, the spin-$U(1)$ symmetry exclude strong spin-orbit coupling. Odd number of rings is preferred to avoid cancelation of  the total Chern number. 
Condition (2) can help to spontaneously break the $\tilde T$ symmetry (where the pattern of the in-plane spin texture due to triplet exciton condensation is not limited to $(p_x+ip_y)$-wave or $(s+id)$-wave), such that QAH insulators can be realized.
Finally, condition (3) can effectively suppress the recombination of particle-excitations and hole-excitations and extend the lifetime of the excitons, given that the interlayer electron-hopping is weaker than the intra-layer coupling.

\begin{figure}[t]
	\includegraphics[width=8.5cm]{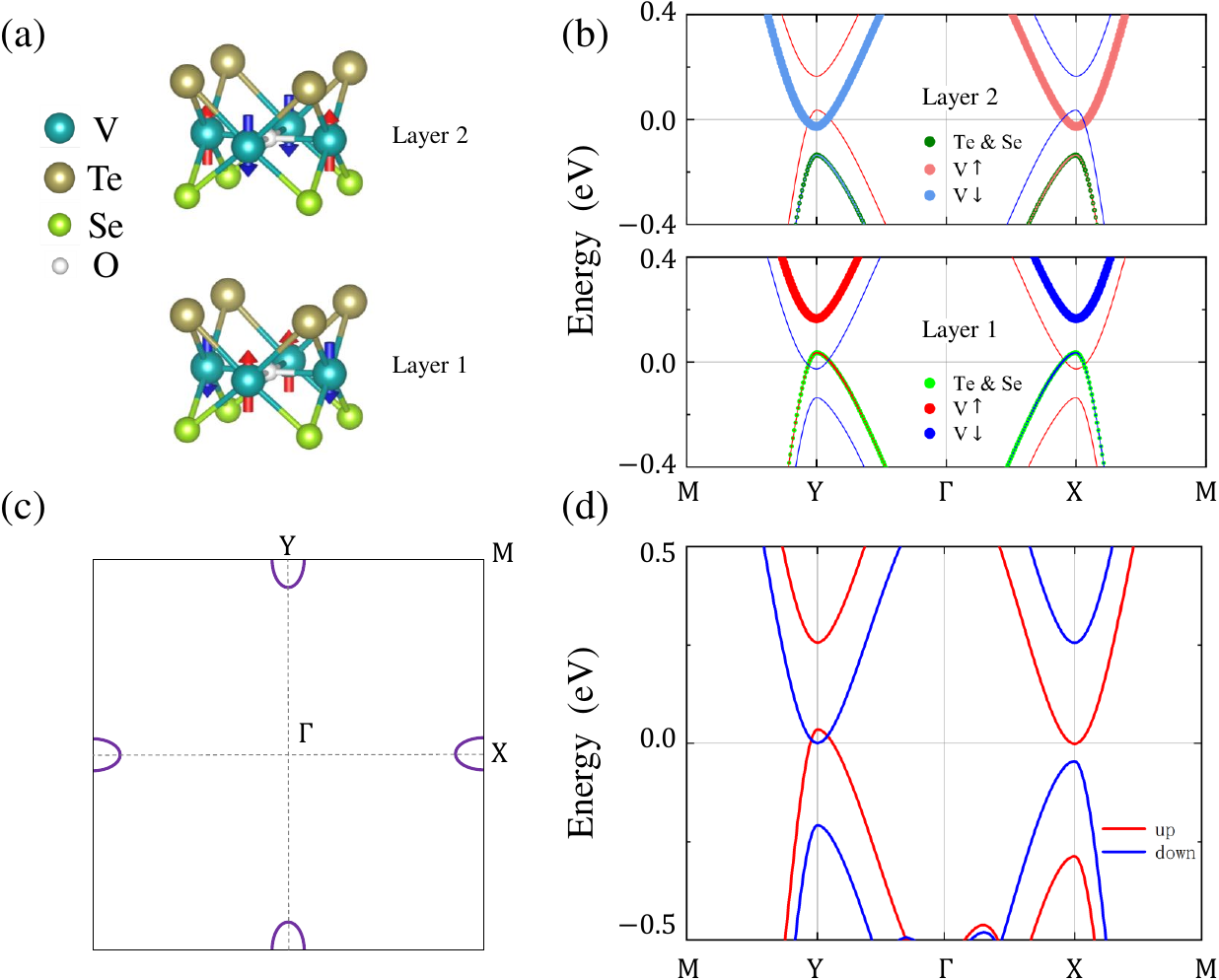}
	\caption{(a) Crystal and magnetic structures of bilayer V$_2$SeTeO. The red and blue arrows represent spin-up and spin-down magnetic moments, respectively.  The displacement between intra-layer $V^{3+}$ ions is 2.86 \AA, the distance between inter-layer $V^{3+}$ ions is 7.5 \AA, and the lattice constant in each layer is 4.05\AA. (b) The electronic structure along the high-symmetry directions and the contributions from different layers without SOC.  (c) The schematic diagram of nodal-line near the Fermi surface for bilayer $\mathrm{V_2SeTeO}$. (d) The electronic structure along the high-symmetry directions when applying uniaxial strain along the y-axis with $b^{\prime}=b(1+0.03)$. }
	\label{materials}
\end{figure}

{\it Candidate materials.} Guided by the above three conditions, here we propose a candidate material, $\mathrm{V_2SeTeO}$\cite{VSeTe} to potentially realize the excitonic QAH insulator. A monolayer $\mathrm{V_2SeTeO}$ has a tetragonal lattice structure (see Fig.\ref{materials} for illustrateion) whose symmetry is described by the symmorphic space group P4mm with the point group $C_{4v}$ generated by the four-fold rotation $C_4^z$ and the mirror reflection $\sigma_x$. It was predicted that a monolayer V$_2$SeTeO is a $d$-wave altermagnetic semiconductor with bandgap 0.15 eV, where the conduction and valence bands have the same spin polarization\cite{VSeTe}. This conclusion is confirmed by our first principle calculations. 

The monolayer of V$_2$SeTeO only contains two magnetic atoms in each unit cell. To reproduce the required band structure of condition (1), one can stack two of monolayers antiferromagnetically (see Fig.\ref{materials}(a)). The band structure is expected to remain $d$-wave altermagnetic, while the spin polarization in the BC band and the TV band are possibly be opposite to each other. Furthermore, the interlayer van-der-Waals interaction may further induce an inversion between the BC and TV bands. Our calculations verified that the bilayer V$_2$SeTeO is indeed a $d$-wave altermagnet with band inversions showing up at the $X$ and $Y$ points (see Fig.\ref{materials}(b)). Moreover,   the BC band and the TV band indeed have opposite spin polarization (see Fig.5(b)\&(d)). Owing to the $U(1)$ symmetry (with spin-orbit coupling ignored), $S_z$ is a good quantum number, so the band inversion result in two nodal rings, rendering the bilayer a nodal-line semimetal  (see Fig.\ref{materials}(c) for the location of the nodal rings) protected by the $U(1)$ symmetry. 
Furthermore, orbital analysis reveals that, the BC band and the TV band are contributed by atoms from different layers (see Fig.\ref{materials}(b)). Since the distance between interlayer atoms are lager than the intra-layer atoms (see Fig.\ref{materials}(a)), the electron-hole recombination will be suppressed and hence the lifetime of the excitons can be considered as very long. Hence, the bilayer V$_2$SeTeO satisfies the conditions for triplet exciton condensation. 

In order to obtain a single nodal-ring structure, we apply uniaxial strain along the y-axis, then only the nodal ring at Y point remains as expected (see Fig.\ref{materials}(d)). Finally, as proposed in \cite{electron-phonon1,electron-phonon2}, long-range magnetic order can enhance the intensity of electron-phonon coupling, which is required to drive the triplet EI into a topological one with QAH effect. Therefore, the material V$_2$SeTeO meets all of the three conditions mentioned above, making it a promising candidate to form triplet excitonic QAH insulator.

{\it Conclusions.}  In summary, we study the lattice models with ferromagnetic and altermagnetic band structure and their triplet exciton condensation. By considering the electron-phonon interaction, we observed the competing between the trivial electron-hole pairing and topological pairing. In this way, we realized QAH excitonic insulators via pure charge-charge interactions without requiring spin-orbit coupling or magnetic flux. We concluded that materials satisfying the following condition are promising to realize a QAH excitonic insulator in 2D: 
(1) odd number of nodal-rings (or odd pairs of nodal-points) at the fermi level protected by $S_z$ conserving symmetry;  
(2) significant electron-phonon coupling and (3) the conducting and valence bands come from different layers. We then proposed the layered material  $\mathrm{V_2SeTeO}$ to be a promising excitonic QAH candidate. 

For simplicity, it is assumed that the attractive interaction is $\pmb q$-independent. We have checked that if $V_{\rm ph}$ is finite for small $q$ but vanishing at large $q$, the main conclusions in the present work still hold valid. On the other hand, we only considered the $Q=0$ exciton condensation and excluded the Fulde-Ferrell-Larkin-Ovchinnikov (FFLO)-type\cite{FF,LO} exciton insulators which spontaneously break translation symmetry. We leave this interesting issue for future study. 

{\it Acknowledgments.} We thank Zhilong Yang, Junwei Liu and especially XiongJun Liu for valuable comments and discussions. This work was supported by the Ministry of Science and Technology of China (No. 2022YFA1405300, No. 2023YFA1406500 and No. 2024YFA1408601) the National Natural Science Foundation of China (Grants No. 12374166, No. 12134020, No.12434009 and No.12204533).
 
\bibliography{references}

@article{altermagnetism-PRX1,
	author = {Libor Šmejkal and Jairo Sinova and Tomas Jungwirth},
	journal = {Phys. Rev. X},
	volume = {12},
	pages = {031042},
	year = {2022},
	url = {https://journals.aps.org/prx/abstract/10.1103/PhysRevX.12.031042}  
}

@article{PRX2,
	author = {Libor Šmejkal and Jairo Sinova and Tomas Jungwirth},
	journal = {Phys. Rev. X},
	volume = {12},
	pages = {040501},
	year = {2022},
	url = {https://journals.aps.org/prx/abstract/10.1103/PhysRevX.12.040501}  
}

@article{PRX3,
	author = {Libor Šmejkal and Anna Birk Hellenes and Rafael González-Hernández and Jairo Sinova and Tomas Jungwirth},
	journal = {Phys. Rev. X},
	volume = {12},
	pages = {011028},
	year = {2022},
	url = {https://journals.aps.org/prx/abstract/10.1103/PhysRevX.12.011028}  
}

@article{SSG1,
	author = {Zhen-Yu Xiao and Jian-Zhou Zhao and Yan-Qi Li and Ryuichi Shindou and Zhi-Da Song},
	journal = {Phys. Rev. X},
	volume = {14},
	pages = {031037},
	year = {2024},
	url = {https://journals.aps.org/prx/abstract/10.1103/PhysRevX.14.031037}  
}

@article{SSG2,
	author = {Xiao-Bing Chen and Jun Ren and Yan-Zhou Zhu and Yu-Tong Yu and Ao Zhang and Peng-Fei Liu and Jia-Yu Li and Yun-Tian Liu and Cai-Heng Li and Qi-Hang Liu},
	journal = {Phys. Rev. X},
	volume = {14},
	pages = {031038},
	year = {2024},
	url = {https://journals.aps.org/prx/abstract/10.1103/PhysRevX.14.031038}  
}

@article{SSG3,
	author = {Yi Jiang and Zi-Yin Song and Tian-Nian Zhu and Zhong Fang and Hong-Ming Weng and Zheng-Xin Liu and Jian Yang and Chen Fang},
	journal = {Phys. Rev. X},
	volume = {14},
	pages = {031039},
	year = {2024},
	url = {https://journals.aps.org/prx/abstract/10.1103/PhysRevX.14.031039}  
}

@article{LiuJunWei,
	author = {Hai-Yang Ma and Meng-Li Hu and Na-Na Li and Jian-Peng Liu and Wang Yao and Jin-Feng Jia and Jun-Wei Liu},
	journal = {Nat Commun},
	volume = {12},
	pages = {2846},
	year = {2021},
	url = {https://doi.org/10.1038/s41467-021-23127-7}  
}

@article{Piezomagnetization1,
	author = {Takuya Aoyama and Kenya Ohgushi},
	journal = {Phys. Rev. Materials},
	volume = {8},
	pages = {L041402},
	year = {2024},
	url = {https://doi.org/10.1103/PhysRevMaterials.8.L041402}  
}

@article{Piezomagnetization2,
	author = {Run-Zhang Xu and Yi-Yan Gao and Jun-Wei Liu},
	journal = {arXiv:},
	volume = {2505},
	pages = {15484},
	year = {2025},
	url = {https://doi.org/10.48550/arXiv.2505.15484}  
}

@article{Piezomagnetization3,
	author = {Yuuki Ogawa and Satoru Hayami},
	journal = {J. Phys. Soc. Jpn.},
	volume = {94},
	pages = {063704},
	year = {2025},
	url = {https://doi.org/10.7566/JPSJ.94.063704}  
}

@article{Piezomagnetization4,
	author = {Chao-Yang Tan and Ze-Feng Gao and Huan-Cheng Yang and  Zheng-Xin Liu and Kai Liu and Peng-Jie Guo and Zhong-Yi Lu},
	journal = {Phys. Rev. B},
	volume = {111},
	pages = {094411},
	year = {2025},
	url = {https://doi.org/10.1103/PhysRevB.111.094411}  
}

@article{Piezomagnetization5,
	author = {Jin-Yang Li and An-Dong Fan and Yong-Kun Wang and Ying Zhang  and Si Li},
	journal = {Appl. Phys. Lett.},
	volume = {125},
	pages = {222404},
	year = {2024},
	url = {https://doi.org/10.1063/5.0242426}  
}

@article{LiuJunWei2,
	author = {Fa-Yuan Zhang and Xing-Kai Cheng and Zhou-Yi Yin and Chang-Chao Liu and Li-Wei Deng and Yu-Xi Qiao and Zheng Shi and Shu-Xuan Zhang and Jun-Hao Lin and Zheng-Tai Liu and Mao Ye and Yao-Bo Huang and Xiang-Yu Meng and Cheng Zhang and Taichi Okuda and Kenya Shimada and Sheng-Tao Cui and Yue Zhao and Guang-Han Cao and Shan Qiao and Jun-Wei Liu and Chao-Yu Chen},
	journal = {Nat. Phys.},
	volume = {21},
	pages = {760–767},
	year = {2025},
	url = {https://doi.org/10.1038/s41567-025-02864-2}  
}

@article{AH1,
	author = {Sajjan Sheoran and Pratibha Dev},
	journal = {Phys. Rev. B},
	volume = {111},
	pages = {184407},
	year = {2025},
	url = {https://doi.org/10.1103/PhysRevB.111.184407}  
}

@article{AH2,
	author = {Toshihiro Sato and Sonia Haddad and Ion Cosma Fulga and Fakher F. Assaad and Jeroen van den Brink},
	journal = {Phys. Rev. Lett.},
	volume = {133},
	pages = {086503},
	year = {2024},
	url = {https://doi.org/10.1103/PhysRevLett.133.086503}  
}

@article{AH3,
	author = {Libor Šmejkal and Allan H. MacDonald and Jairo Sinova and Satoru Nakatsuji and Tomas Jungwirth},
	journal = {Nat Rev Mater},
	volume = {7},
	pages = {482–496},
	year = {2022},
	url = {https://doi.org/10.1038/s41578-022-00430-3}  
}

@article{AH4,
	author = {Ze-Xin Feng and Xiao-Rong Zhou and Libor Šmejkal and Lei Wu and Zeng-Wei Zhu and Hui-Xin Guo and Rafael González-Hernández and XiaoNing Wang and Han Yan and Pei-Xin Qin and Xin Zhang and Hao-Jiang Wu and Hong-Yu Chen and Ziang Meng and Li Liu and Zheng-Cai Xia and Jairo Sinova and Tomáš Jungwirth and Zhiqi Liu},
	journal = {Nat Electron},
	volume = {5},
	pages = {735–743},
	year = {2022},
	url = {https://doi.org/10.1038/s41928-022-00866-z}  
}

@article{AH5,
	author = {Miina Leiviskä and Javier Rial and Antonín Bad'ura and Rafael Lopes Seeger and Ismaïla Kounta and Sebastian Beckert and Dominik Kriegner and Isabelle Joumard and Eva Schmoranzerová and Jairo Sinova and Olena Gomonay and Andy Thomas and Sebastian T. B. Goennenwein and Helena Reichlová and Libor Šmejkal and Lisa Michez and Tomáš Jungwirth and Vincent Baltz},
	journal = {Phys. Rev. B},
	volume = {109},
	pages = {224430},
	year = {2024},
	url = {https://doi.org/10.1103/PhysRevB.109.224430}  
}

@article{AH6,
	author = {Libor Šmejkal and Rafael González-Hernández and T. Jungwirth and J. Sinova},
	journal = {Sci. Adv},
	volume = {6},
	pages = {eaaz8809},
	year = {2020},
	url = {10.1126/sciadv.aaz8809}  
}

@article{Haldane,
	author = {F. D. M. Haldane},
	journal = {Phys. Rev. Lett.},
	volume = {61},
	pages = {2015},
	year = {1988},
	url = {https://journals.aps.org/prl/abstract/10.1103/PhysRevLett.61.2015}  
}

@article{LIuZhangQi_ARCMP16,
	author = {C.-X. Liu and S.-C. Zhang and X.-L. Qi},
	journal = {Annu. Rev. Conden. Ma. P},
	volume = {7},
	pages = {301},
	year = {2016},
	url = {https://www.annualreviews.org/content/journals/10.1146/annurev-conmatphys-031115-011417}  
}

@article{QiWuZhang,
	author = {X.-L. Qi and Y.-S. Wu and S.-C. Zhang},
	journal = {Phys. Rev. B},
	volume = {74},
	pages = {085308},
	year = {1988},
	url = {https://journals.aps.org/prb/abstract/10.1103/PhysRevB.74.085308}  
}

@article{GuoLiuLu_npj23,
	author = {P.-J. Guo and Z.-X. Liu and Z.-Y. Lu},
	journal = {npj Comput Mater},
	volume = {9},
	pages = {70},
	year = {2023},
	url = {https://www.nature.com/articles/s41524-023-01025-4}  
}

@article{WuShanYan_PRL14,
	author = {Shu-Chun Wu and Guang-Cun Shan and Bing-Hai Yan},
	journal = {Phys. Rev. Lett.},
	volume = {113},
	pages = {256401},
	year = {2014},
	url = {https://journals.aps.org/prl/abstract/10.1103/PhysRevLett.113.256401}  
}

@article{Xue_Science13,
	author = {C.-Z. Chang and J.-S. Zhang and X. Feng and J. Shen and Z.-C. Zhang and M.-H. Guo and K. Li and Y.-B. Ou and P. Wei and L.-L. Wang and Z.-Q. Ji and Y. Feng and S.-H. Ji and X. Chen and J.-F. Jia and X. Dai, Z. Fang and S.-C. Zhang and K. He and Y.-Y. Wang and L. Lu and X.-C. Ma and Q.-K. Xue},
	journal = {Science},
	volume = {340},
	pages = {167},
	year = {2013},
	url = {https://www.science.org/doi/10.1126/science.1234414}  
}

@article{SM1,
	author = {Cong Li and Meng-Li Hu and Zhi-Lin Li and Yang Wang and Wan-Yu Chen and Balasubramanian Thiagarajan and Mats Leandersson and Craig Polley and Timur Kim and Hui Liu and Cosma Fulga and Maia G. Vergniory and Oleg Janson and Oscar Tjernberg and Jeroen van den Brink},
	journal = {Commun Phys},
	volume = {8},
	pages = {311},
	year = {2025},
	url = {https://doi.org/10.1038/s42005-025-02232-9}  
}

@article{SM2,
	author = {Wen-Long Lu and Shi-Yu Feng and Yu-Zhi Wang and Dong Chen and Zi-Han Lin and Xin Liang and Si-Yuan Liu and Wan-Xiang Feng and Kohei Yamagami and Jun-Wei Liu and Claudia Felser and Quan-Sheng Wu and Junzhang Ma},
	journal = {Nano Lett.},
	volume = {25, 18},
	pages = {7343},
	year = {2025},
	url = {https://pubs.acs.org/doi/10.1021/acs.nanolett.5c00482}  
}

@article{SM3,
	author = {Shuai Qu and Xiao-Yao Hou and Zheng-Xin Liu and Peng-Jie Guo and Zhong-Yi Lu},
	journal = {Phys. Rev. B},
	volume = {111},
	pages = {195138},
	year = {2025},
	url = {https://doi.org/10.1103/PhysRevB.111.195138}  
}

@article{SM4,
	author = {Chao-Yang Tan and Ze-Feng Gao and Huan-Cheng Yang and Kai Liu and Peng-Jie Guo and Zhong-Yi Lu},
	journal = {arXiv:},
	volume = {2406},
	pages = {16603},
	year = {2024},
	url = {https://doi.org/10.48550/arXiv.2406.16603}  
}

@article{TI1,
	author = {Hai-Yang Ma and Jin-Feng Jia},
	journal = {Phys. Rev. B},
	volume = {110},
	pages = {064426},
	year = {2024},
	url = {https://doi.org/10.1103/PhysRevB.110.064426}  
}

@article{TI2,
	author = {Zheng-Tian Li and Ze-Yu Li and and Zhen-Hua Qiao},
	journal = {Phys. Rev. B},
	volume = {111},
	pages = {155303},
	year = {2025},
	url = {https://doi.org/10.1103/PhysRevB.111.155303}  
}

@article{TI3,
	author = {Rafael González-Hernández and Higinio Serrano and Bernardo Uribe},
	journal = {Phys. Rev. B},
	volume = {111},
	pages = {085127},
	year = {2025},
	url = {https://doi.org/10.1103/PhysRevB.111.085127}  
}

@article{TSC1,
	author = {Di Zhu and Zheng-Yang Zhuang and Zhi-Gang Wu and Zhong-Bo Yan},
	journal = {Phys. Rev. B},
	volume = {108},
	pages = {184505},
	year = {2023},
	url = {https://doi.org/10.1103/PhysRevB.108.184505}  
}

@article{TSC2,
	author = {Seung Beom Hong and Moon Jip Park and Kyoung-Min Kim},
	journal = {Phys. Rev. B},
	volume = {111},
	pages = {054501},
	year = {2025},
	url = {https://doi.org/10.1103/PhysRevB.111.054501}  
}

@article{TSC3,
	author = {Sayed Ali Akbar Ghorashi and Taylor L. Hughes and Jennifer Cano},
	journal = {Phys. Rev. Lett.},
	volume = {133},
	pages = {106601},
	year = {2024},
	url = {https://doi.org/10.1103/PhysRevLett.133.106601}  
}

@article{EC1,
	author = {John M. Blatt and K. W. Böer and Werner Brandt},
	journal = {Phys. Rev.},
	volume = {126},
	pages = {1691},
	year = {1962},
	url = {https://doi.org/10.1103/PhysRev.126.1691}  
}

@article{EC2,
	author = {D. Jérome and T. M. Rice and W. Kohn},
	journal = {Phys. Rev.},
	volume = {158},
	pages = {462},
	year = {1967},
	url = {https://doi.org/10.1103/PhysRev.158.462}  
}

@article{MacDonald_PRB,
	author = {Y.-P. Shim and A. H. MacDonald},
	journal = {Phys. Rev. B},
	volume = {79},
	pages = {235329},
	year = {2009},
	url = {https://doi.org/10.1103/PhysRevB.79.235329}
}

@article{Metal2,
	author = {Qiang Gao and Yang-hao Chan and Yu-Zhe Wang and Hao-Tian Zhang and Jin-Xu Pu and Sheng-Tao Cui and Yi-Chen Yang and Zheng-Tai Liu and Da-Wei Shen and Zhe Sun and Juan Jiang and Tai-C. Chiang and Peng Chen},
	journal = {Nat Commun},
	volume = {14},
	pages = {994},
	year = {2023},
	url = {https://doi.org/10.1038/s41467-023-36667-x}  
}

@article{TEI1,
	author = {Zhi-Huan Dong and Ya-Hui Zhang},
	journal = {Phys. Rev. B},
	volume = {107},
	pages = {L081101},
	year = {2023},
	url = {https://doi.org/10.1103/PhysRevB.107.L081101}  
}

@article{TEI2,
	author = {Ying-Ming Xie and Cheng-Ping Zhang and K. T. Law},
	journal = {Phys. Rev. B},
	volume = {110},
	pages = {045115},
	year = {2024},
	url = {https://doi.org/10.1103/PhysRevB.110.045115}  
}

@article{spintex,
	author = {J. Kuneš and D. Geffroy},
	journal = {Phys. Rev. Lett.},
	volume = {116},
	pages = {256403},
	year = {2016},
	url = {https://doi.org/10.1103/PhysRevLett.116.256403}  
}

@article{spintriplet,
	author = {Ze-Yu Jiang and Wen-Kai Lou and Yu Liu and Yuan-Chang Li and Hai-Feng Song and Kai Chang and Wen-Hui Duan and Sheng-Bai Zhang},
	journal = {Phys. Rev. Lett.},
	volume = {124},
	pages = {166401},
	year = {2020},
	url = {https://doi.org/10.1103/PhysRevLett.124.166401}  
}

@article{VSeTe,
	author = {Wei Xun and Xin Liu and You-Dong Zhang and and Yin-Zhong Wu and Ping Li},
	journal = {Appl. Phys. Lett.},
	volume = {126},
	pages = {161903},
	year = {2025},
	url = {https://doi.org/10.1063/5.0267525}  
}

@article{Column,
	author = {Yves H. Kwan and T. Devakul and S. L. Sondhi and S. A. Parameswaran},
	journal = {Phys. Rev. B},
	volume = {104},
	pages = {125133},
	year = {2021},
	url = {https://doi.org/10.1103/PhysRevB.104.125133}  
}

@article{Column2,
	author = {Yan-De Que and Yang-Hao Chan and Jun-Xiang Jia and Anirban Das and Zheng-Jue Tong and Yu-Tzu Chang and Zhen-Hao Cui and Amit Kumar and Gagandeep Singh and Shantanu Mukherjee and Hsin Lin and Bent Weber},
	journal = {Adv. Mater.},
	volume = {36},
	pages = {2309356},
	year = {2024},
	url = {https://doi.org/10.1002/adma.202309356}  
}

@article{electron-phonon1,
	author = {L. Boeri and M. Calandra and I. I. Mazin and O. V. Dolgov and and F. Mauri},
	journal = {Phys. Rev. B},
	volume = {82},
	pages = {020506(R)},
	year = {2010},
	url = {https://doi.org/10.1103/PhysRevB.82.020506}  
}

@article{electron-phonon2,
	author = {Rui-Qi Zhang and Yan-Yong Wang and Manuel Engel and Christopher Lane and Henrique Miranda and Lin Hou and Sugata Chowdhury and Bahadur Singh and Bernardo Barbiellini and Jian-Xin Zhu and Robert S. Markiewicz and E. K. U. Gross and Georg Kresse and Arun Bansil and Jian-Wei Sun},
	journal = {arXiv},
	volume = {2504},
	pages = {13025},
	year = {2025},
	url = {https://doi.org/10.48550/arXiv.2504.13025}  
}

@article{FF,
	author = {Peter Fulde and Richard A. Ferrell},
	journal = {Phys. Rev.},
	volume = {135},
	pages = {A550},
	year = {1964},
	url = {https://doi.org/10.1103/PhysRev.135.A550}  
}

@article{LO,
	author = {A. I. Larkin and Y. N. Ovchinnikov},
	journal = {Sov. Phys. JETP},
	volume = {20},
	pages = {726},
	year = {1965},
}

@article{YangLiuFang21B,
  title = {Symmetry-protected nodal points and nodal lines in magnetic materials},
  author = {Yang, Jian and Fang, Chen and Liu, Zheng-Xin},
  journal = {Phys. Rev. B},
  volume = {103},
  issue = {24},
  pages = {245141},
  numpages = {56},
  year = {2021},
  month = {Jun},
  publisher = {American Physical Society},
  doi = {10.1103/PhysRevB.103.245141},
  url = {https://link.aps.org/doi/10.1103/PhysRevB.103.245141}
}

@article{YangLiuFang24NC,
	author = {Yang, Jian and Liu, Zheng-Xin and Fang, Chen},
	da = {2024/11/25},
	doi = {10.1038/s41467-024-53862-6},
	id = {Yang2024},
	isbn = {2041-1723},
	journal = {Nature Communications},
	number = {1},
	pages = {10203},
	title = {Symmetry invariants and classes of quasiparticles in magnetically ordered systems having weak spin-orbit coupling},
	ty = {JOUR},
	url = {https://doi.org/10.1038/s41467-024-53862-6},
	volume = {15},
	year = {2024}}

@article{SongZD25L,
  title = {Intrinsic Axion Statistical Topological Insulator},
  author = {Chen, Xi and Wang, Fa-Jie and Bi, Zhen and Song, Zhi-Da},
  journal = {Phys. Rev. Lett.},
  volume = {134},
  issue = {22},
  pages = {226601},
  numpages = {8},
  year = {2025},
  month = {Jun},
  publisher = {American Physical Society},
  doi = {10.1103/PhysRevLett.134.226601},
  url = {https://link.aps.org/doi/10.1103/PhysRevLett.134.226601}
}

@article{Laughlin_argu,
  title = {Quantized Hall conductivity in two dimensions},
  author = {Laughlin, R. B.},
  journal = {Phys. Rev. B},
  volume = {23},
  issue = {10},
  pages = {5632--5633},
  numpages = {0},
  year = {1981},
  month = {May},
  publisher = {American Physical Society},
  doi = {10.1103/PhysRevB.23.5632},
  url = {https://link.aps.org/doi/10.1103/PhysRevB.23.5632}
}

\newpage

\section*{End Matter}

\appendix
{\it Mean-field equations and the ground state energy}
The self-consistent mean field equations for the parameters $ \Delta_{\pmb{k}}$ are given by:
\begin{eqnarray} \label{MFeq3}
	\Delta_{\pmb{k}} &=& \frac{1}{2N}\sum_{\pmb{k}^{'}} U(\pmb{k}-\pmb{k}^{'})
	 F_{\pmb{k}^{'},\pmb{k}}^{cc}F_{\pmb{k},\pmb{k}^{'}}^{vv}  \dfrac{\Delta_{\pmb{k}}}{\sqrt{\epsilon_{\pmb{k}}^{2}+|\Delta_{\pmb{k}}|^{2}}}. 
\end{eqnarray}
Here we only consider the interband interactions which are essential for the exciton condensation. The intraband interactions only affect the diagonal part $\epsilon_{c(v)\pmb k}^0$ and are ignored in our discussion. The form factors $F_{\pmb{k},\pmb{k}^{'}}^{dd^{'}}
=\sum_\beta \varphi_{\beta,d\pmb k}^*\varphi_{\beta, d'\pmb k'}$ 
can be calculated from the eigenstates $\varphi_{\beta, d\pmb k}$ of $H_0$ given in the main text.

To obain the phase diagram, we need to determine which state has the lowest energy for given interaction parameters. Depending on the pairing symmetry of the initial values of $\Delta_{\pmb{k}}$, one can calculate the energy of each trial state
\begin{eqnarray}
	E_{\rm trial}&=&\frac{1}{N}\left( E_{0}+\sum_{\pmb{k}}E_{\pmb{k}}^{-}\right), 
\end{eqnarray}
where $E_{\pmb{k}}^{-}$ is given in the main text and $E_0$ is given by
\begin{eqnarray}
	E_0&=&\frac{1}{2}\sum_{\pmb{k}} \Delta_{\pmb{k}} \left\langle b_{v\pmb{k}}^\dag b_{c\pmb{k}}\right\rangle+\frac{1}{2}\sum_{\pmb{k}} \Delta_{\pmb{k}}^* \left\langle b_{c\pmb{k}}^\dag b_{v\pmb{k}}\right\rangle .
\end{eqnarray}
In the present study, we compare the solutions with topological exciton condensation ($(p_x+ip_y)$-wave or $(s+id)$-wave), trivial exciton condensate ($p$-wave) and metalic state without exciton condensation. The state having the lowest trial energy is considered to be the ground state. 

{\it Spin texture and the `pairing symmetry' of excitons} 
Due to the condensation of spin triplet exciton, the electrons exhibit in-plane polarization at each $\pmb k$ point. The total magnetization is given by $$M_{x,y} = \frac{1}{N}\sum_{\pmb k} \langle S^{x,y}(\pmb{k})\rangle,$$ where

\begin{eqnarray}\label{Eq22}
\left\langle S^x(\pmb{k})\right\rangle &=& \frac{1}{2}\sum_{\alpha} \left( \left\langle c_{\alpha,k\up}^\dag c_{\alpha,k\dn}\right\rangle + \left\langle c_{\alpha,k\dn}^\dag c_{\alpha,k\up}\right\rangle\right),\notag\\
	\left\langle S^y(\pmb{k})\right\rangle &=& -\frac{i}{2} \sum_{\alpha} \left( \left\langle c_{\alpha,k\up}^\dag c_{\alpha,k\dn}\right\rangle - \left\langle c_{\alpha,k\dn}^\dag c_{\alpha,k\up}\right\rangle\right), \notag
\end{eqnarray}
and 
\Beq
\left\langle c_{\alpha,k\up}^\dag c_{\alpha,k\dn}\right\rangle = \sum_{nn^{'}=cv} \varphi^{*}_{{2\alpha-1}, n\pmb{k}} \varphi_{{2\alpha}, n^{'}\pmb{k}} \left\langle b_{n\pmb{k}}^\dag b_{n^{'}\pmb{k}}\right\rangle. \notag
\Eeq

Since the phases of the eigenstates $\varphi_{\beta,d \pmb k}$ have ambiguity due to the gauge degrees of freedom, the phases of the order parameters $\Delta_{\pmb k}$ of the exciton condensation are gauge dependent. However, the expectation value of the spin operators $\langle S^{x,y}(\pmb{k})\rangle$ are independent on the gauge choices. Therefore, we use the $\pmb k$ dependence of $\langle S^{x}(\pmb{k})\rangle$, $\langle S^{y}(\pmb{k})\rangle$, namely the spin-texture in momentum space, to define the `pairing symmetry' of the spin triplet exciton condensation.  For instance, the spin texture in Fig.\ref{fig:FM-EC}(b) is $d$-wave while the one in Fig.\ref{fig:FM-EC}(e) is $(p_x+ip_y)$-wave like. The pairing symmetry of topological exciton condensate in the AM model is $(s+id)$-wave like, as shown in Fig.\ref{AM-EC}(e).

Noticing that the total in-plane magnetization contributed by sublattice-$1$ and sublattice-$2$ exactly cancel, 
when judging the exciton pairing symmetry from  $\langle S^{x}(\pmb{k})\rangle$ and $\langle S^{y}(\pmb{k})\rangle$, we only consider the contribution from the sublattice-$1$. 

\begin{figure}[b]
	\includegraphics[width=8cm]{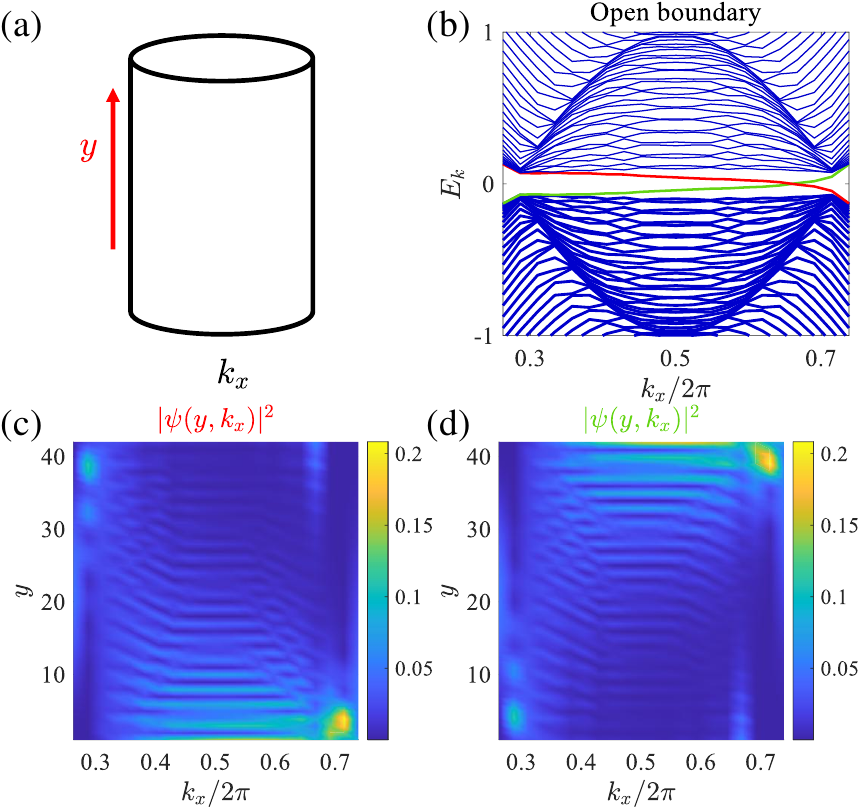}
	\caption{The gapless chiral states of Topological excitonic insulator for ferromagnetism in the main text. (a) A geometry with the open boundary condition. (b) The energy spectrum with two gapless edge modes with (c) the red mode localized on the boundary $y=40$ and (d) the green mode localized on the boundary $y=1$.}
	\label{Wf}
\end{figure}

{\it The chiral edge states}
Here we show the existence of chiral edge modes for the $(p_x+ip_y)$-wave exciton condensed insulator in the FM model which has nonzero Chern number. For a system with open boundary along $y$-direction and periodic boundary condition along $x$-direction (see Fig.\ref{Wf}(a)), the energy spectrum is shown in Fig.\ref{Wf}(b), where two interband chiral modes manifest themselves linking the conduction and valence bands. 

To illustrate that the chiral modes are indeed locating on the edges, we plot $|\psi(k_x,y)|^2$ as a function of $k_x$ and $y$. As shown in Fig.\ref{Wf}(c)\&(d), the left-moving chiral state is locating on the bottom edge, and the right-moving chiral state is locating on the top edge. The existence of these gapless chiral edge states verifies the quantization of Hall conductance according to Laughlin's argument\cite{Laughlin_argu}. 

\begin{figure}[t]
	\includegraphics[height=6cm]{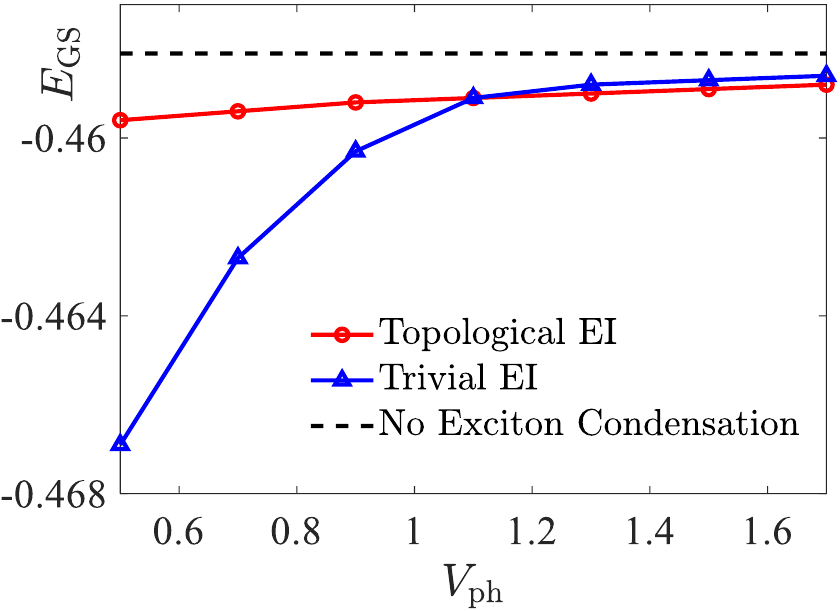}
	\caption{The total ground
		energy $E_{\rm GS}$ of topological EIs, trivial EIs and metals (no exciton condensation) with various $V_{\rm ph}$ for AM model with $g=1$.}
	\label{AMphase}
\end{figure}

{\it Exciton condensation in altermagnetic mode}\ The altermagnetic mode we are considering contains four-sublattices labeled by 1,2,3,4. The tight-binding Hamiltonian of the electrons on the lattice then reads,
\begin{eqnarray}
		H_0^{(AM)}&=& \sum_{j}  t_{1} \left(\sum_{\alpha=1,2}\sum_{\alpha^{\prime}=3,4}C_{\alpha, j}^\dag C_{\alpha^{\prime}, j}+\text { h.c. }\right)  \nonumber \\ 
		& &+\sum_{j}  t_{3} \left(\sum_{d_1=\pmb{x},\pmb{y}}(C_{1,j}^\dag C_{3,j+d_1}+C_{2,j}^\dag C_{4,j-d_1}) \right. \nonumber \\ 
		&&\left. +\sum_{d_2=\pmb{x},\pmb{-y}}(C_{1,j}^\dag C_{4,j+d_2}+C_{2,j}^\dag C_{3,j-d_2})+\text { h.c. }\right)  \nonumber \\ 
		& &+\sum_{j} \left(t_2^x (C_{1,j}^\dag C_{2, j}+C_{1,j}^\dag C_{2, j-\pmb x})\right. \nonumber \\
		&&\left. +t_2^y (C_{3, j}^\dag C_{4, j}+C_{3, j}^\dag C_{4, j+\pmb y})+\text { h.c. }\right) \nonumber \\ 
		& &+\sum_{\alpha, j}\left( t_4^x C_{\alpha, j}^\dag C_{\alpha, j\pm \pmb x}+t_4^y C_{\alpha, j}^\dag C_{\alpha, j\pm \pmb y}\right)\nonumber \\ 
		& &+\sum_{\alpha, j} C_{\alpha, j}^\dag \left(\mu_{\alpha} \sigma_{0}+\pmb{m}_{\alpha} \cdot {\pmb{\sigma}\over2} \right) C_{\alpha, j}.	\label{AMEq1}
\end{eqnarray}

When considering electron-electron interactions, there are also two exciton insulators with different pairing symmetries similar to the Ferromagnetic model. As shown in Fig.\ref{AMphase}, the electron-phonon coupling can induce the phase transition from the trivial EI ($p$-wave like) and the topological EI ($(s+id)$-wave like). Since the nodal-ring is centered at the Y point, the energies of the trivial EI and topological EI are both dependent on $V_{\rm ph}$, which is different from the FM model in Fig.\ref{fig:Transition}(b).

 \begin{figure}[t]
	\includegraphics[width=8.6cm]{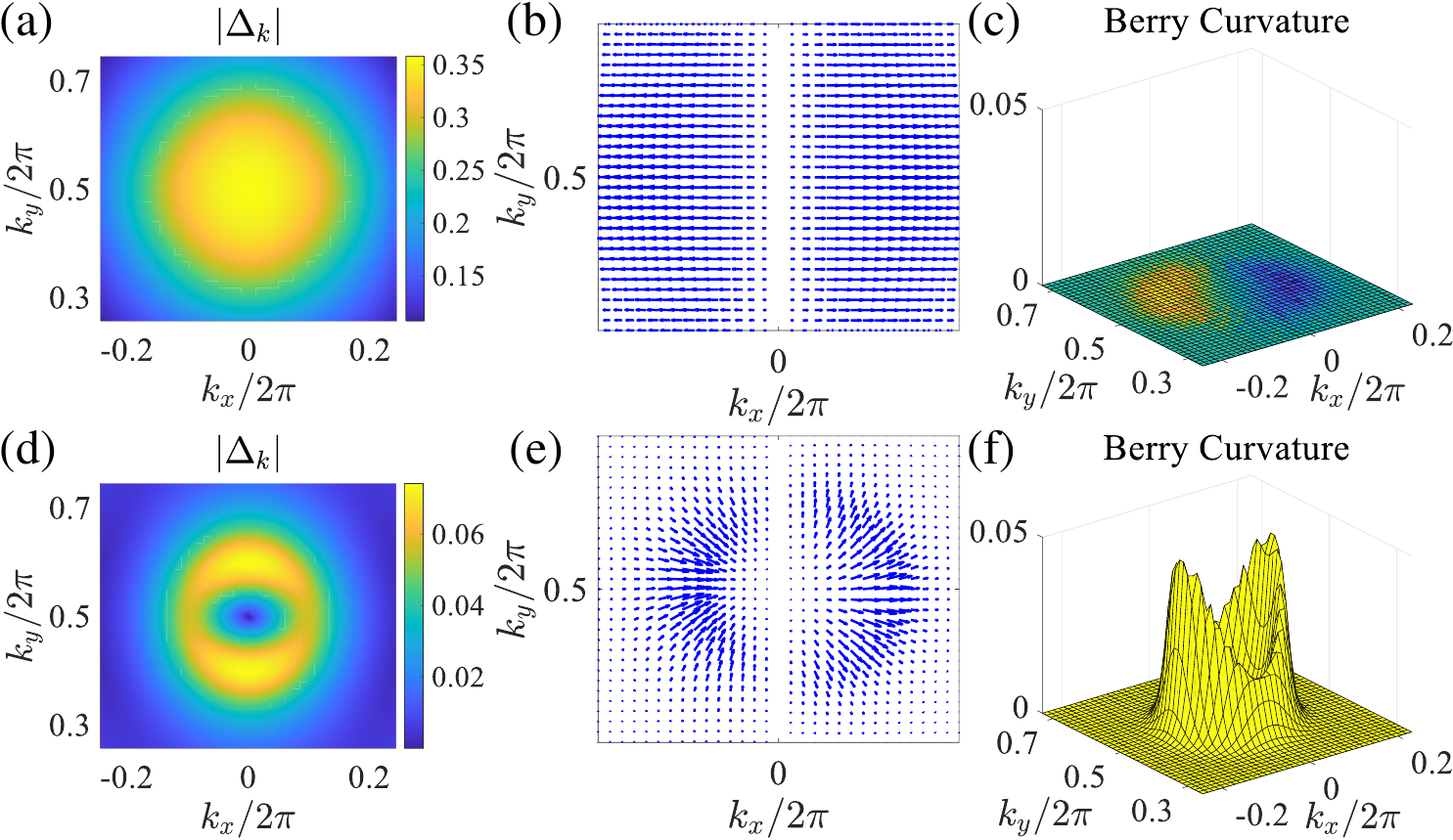}
	\caption{ The results of exciton condensation in altermagnetism with a nodal ring at Y point  for $g=1$. (a) Magnitude of the order parameters $|\Delta_{\pmb{k}}|$, (b) the spontaneous $p$-wave collinear spin textures and (c) the berry curvature for trivial EI with $C=0$ for $V_{\rm ph}=0$. (d) Magnitude of the order parameters $|\Delta_{\pmb{k}}|$, (e) the spontaneous $(s+id)$-wave noncollinear spin textures and (f) the berry curvature for topological EI with $|C|=1$ for $V_{\rm ph}=1.3$.}
	\label{AM-EC}
\end{figure}

Here we analyze the band structure of exciton insulators in the AM model (\ref{AMEq1}). With only Coulomb  interactions for $V_{\rm ph}=0$, a trivial EI with Chern number $C=0$ is energetically favored. As shown in Fig.\ref{AM-EC}(b), the texture of the in-plane magnetization shows a $p$-wave collinear pattern in the BZ with zero net in-plane magnetization. The magnitude of the exciton condensate is shown in Fig.\ref{AM-EC}(a). Due to the remaining effective time reversal symmetry $\tilde T\equiv \left(C_{2}^{\pmb n}T||T\right)$ ($\pmb n$ is perpendicular to the direction of the magnetic momentum), the Berry curvature as shown in Fig.\ref{AM-EC}(c) is zero.

When turning on the electron-phonon coupling by setting $V_{\rm ph}=1.3$, a topological EI with $|C|=1$ is obtained. The magnitude of the exciton condensate is shown in Fig.\ref{AM-EC}(d), which contains a node at the Y point.  Since the in-plane magnetization exhibits a $(s+id)$-wave noncollinear pattern in the BZ [see Fig.\ref{AM-EC}(e)], the net in-plane magnetization $M_x$ is nonzero. This is an observable quantity to detect the exciton condensation. The noncollinear magnetic structure yields nonzero Berry curvature [see Fig.\ref{AM-EC}(f)] with Chern number $|C|=1$. Hence the excitonic anomalous quantum Hall insulator is realized via altermagnetism. 
\newpage

\end{document}